\documentstyle[12pt,epsf,epsfig,wrapfig]{article}
\begin{document}
\begin{center}
{\bfseries An Alternative Approach to Semi-inclusive DIS
\\and A Model Independent Determination of $D_u^{\pi^+}$, $D_d^{\pi^+}$ and $D_s^{\pi^+}$ }
\footnote{ A talk given by E. Christova at IX Workshop on High Energy Physics, SPIN 2001,
Dubna, Russia}
\vskip 5mm
E. Christova$^{1 \dag}$, E. Leader$^{2}$ and S.
Kretzer$^{3}$
 \vskip 5mm
 {\small (1) {\it Institute of Nucl. Res.
and Nuclear Energy, Sofia }
\\
(2) {\it Imperial College, London }
\\
(3) {\it Michigan University, Michigan }
 \\
 $\dag$ {\it E-mail:
echristo@inrne.bas.bg, e.leader@ic.ac.uk, kretzer@pa.msu.edu }}
\end{center}
\vskip 5mm

\begin{abstract}
\begin{center}
We discuss how to extract the polarized parton
densities  and fragmentation functions  in a model
independent way from SIDIS,
 and present results for $D_{u,d,s}^{\pi^+}$.
\end{center}
\end{abstract}

\vskip 10mm

There are two types of processes that give information on
polarized parton densities in  nucleon. The polarized asymmetry
in DIS of longitudinally polarized leptons on longitudinally
polarized nucleons:
\begin{eqnarray}
\Delta A_N^{DIS}
=\frac{2y(2-y)}{1+(1-y^2)}\frac{\Delta
\sigma_N^{DIS}}{\sigma_N^{DIS}} =\frac{ \sum_{q,\bar q} e^2_q
\,\Delta q_i(x,Q^2)}{\sum_{q,\bar q} e^2_q \,q_i(x,Q^2)}\qquad
\left(LO\right)
 \end{eqnarray}
measures (both in LO and NLO in QCD) the combinations:
\begin{eqnarray}
\Delta
u+\Delta\bar u, \quad \Delta d+\Delta\bar d,\quad \Delta
s+\Delta\bar s, \quad \Delta G.\label{deltaq}
 \end{eqnarray}
Thus in principle, we cannot determine separately the valence and sea quark
distributions, independently of the precision and
the amount of data available.

The polarized asymmetry of semi-inclusive DIS (SIDIS) of
longitudinally polarized leptons on longitudinally polarized nucleons
\begin{eqnarray}
 \overrightarrow{e} + \overrightarrow{N}\to e+ h+ X,\label{theprocess}
\end{eqnarray}
when a final hadron $h$ is detected
\begin{eqnarray}
\Delta A_N^h=\frac{2y(2-y)}{1+(1-y^2)}
\frac{\Delta \sigma_N^h}{\sigma_N^h}=\frac{\sum_{q,\bar q} e^2_q\,
\Delta q_i(x,Q^2)\,
D_i^h(z,Q^2)}{\sum_{q,\bar q} e^2_q \,q_i(x,Q^2)\,
D_i^h(z,Q^2)}\qquad \left(LO\right)
\end{eqnarray}
has the advantage that it measures the $\Delta q$ and
$\Delta\bar q$ separately. However,
a knowledge of the fragmentation functions (FFs) is required.
The $D_q^h$
are supposed to be known from the inclusive process $e^+e^-\to h + X$, $h=\pi
,K, p,...$:
 \begin{eqnarray}
 \sigma^h \propto \sum e_i^2 (D_q^h + D_{\bar q}^h). \qquad \left(LO\right)
\end{eqnarray}
In principle, similar to DIS (both in LO and NLO),
only the combination $D_q^h +D_{\bar q}^h$ can be determined.
 However, for
the interpretation of SIDIS data, $D_q^h$ and $D_{\bar q}^h$ are
needed separately. Usually additional theoretical assumptions
about favoured and unfavoured transitions are made. This leads to
rather different parametrizations of the FFs obtained from
analyzing the $e^+e^-$ data~\cite{FFparams}. In addition, in order
to reduce the parameters, different assumptions about the
polarized sea densities are made.

In this talk we consider what we can learn from SIDIS - both with
polarized and unpolarized initial particles - without assuming any
 knowledge about FFs and without any assumptions about the polarized sea
 $\Delta \bar q$. We shall use information that comes directly from
  {\it measurable quantities} in the {\it same experiment}.
  A somewhat similar approach has been considered by S.~Manayenkov~\cite{Manaenko}.

  \vspace{1cm}

 {\bf 2.} As known quantities in our approach we consider the unpolarized
 parton densities and the polarized non-singlet combination $\Delta q_3=
 (\Delta u + \Delta\bar u)-(\Delta d + \Delta\bar d)$ that is determined
 directly from DIS:
 \begin{eqnarray}
g_1^p(x,Q^2) -g_1^n(x,Q^2) =\frac{1}{6} \Delta q_3\otimes \left(1+
\frac{\alpha_s(Q^2)}{2\pi}\delta C_q\right)\label{deltaq3}
\end{eqnarray}
either in LO or NLO. For $\Delta q_3$  we do not use its value obtained from
the parametrizations  of
$(\Delta u + \Delta\bar u)$ and $(\Delta d + \Delta\bar d)$ since the latter
are influenced by the less known quantities $\Delta s$ and $\Delta G$.

We work with  three measurable quantities: the polarized asymmetries of DIS
 and SIDIS, $\Delta A_N^{DIS}$ and $\Delta A_N^{h}$, and the ratio
\begin{eqnarray}
R_N^h=\frac{\sigma^h_N}{\sigma_N^{DIS}}
\end{eqnarray}
of the unpolarized SIDIS normalized to the inclusive DIS cross section.

\vspace{1cm}

{\bf 3.} In LO the polarized valence quark densities can
 be obtained, without any knowledge of the FFs assuming only isospin invariance,
from simple algebraic equations. For $\pi^\pm$ production we have:
\begin{eqnarray}
\Delta u_V (x,Q^2)&=& \frac{1}{15}\left\{4(4u_V - d_V)\Delta A_{p}^{\pi^+-
\pi^-}
+ (4d_V -u_V)\Delta A_{n}^{\pi^+-\pi^-}\right\}\nonumber\\
\Delta d_V (x,Q^2)&=&\frac{1}{15}\left\{4(4d_V - u_V)\Delta A_{n}^{\pi^+-
\pi^-}
+ (4u_V - d_V)\Delta A_{p}^{\pi^+-\pi^-}\right\}.\label{A}
\end{eqnarray}
Analogously, $\Delta u_V $ and $\Delta d_V $ can be obtained also from
 $K^\pm$ and $\Lambda,\bar\Lambda$ production~\cite{Strategy}.

 Once we know the valence quark densities, we can  determine the
 SU(2) breaking of the polarized sea, without requiring any knowledge of
 the unknown $\Delta\bar q$ and $\Delta G$. We have:
\begin{eqnarray}
[\Delta \bar u(x,Q^2)-\Delta \bar d(x,Q^2)]_{LO} =
\frac{1}{2}\left[\Delta q_3(x,Q^2) + \Delta  d_V(x,Q^2)-
\Delta  u_V(x,Q^2)\right]_{LO}.\label{sea}
\end{eqnarray}

Given that $u_V(x,Q^2)$ and $d_V(x,Q^2)$ are well
determined from inclusive DIS one can proceed further to obtain
the non-singlet combinations of FF's $D_q^h - D_q^{\bar h} \equiv
D_q^{h-\bar h}$ from unpolarized SIDIS.
For $\pi^\pm$ we have
 \begin{eqnarray}
D_u^{\pi^+ -\pi^- }(z,Q^2) =\frac{18\left( F_{1}^p\right)_{LO} R_p^{\pi^+
-\pi^-}}{4u_V-d_V}, \quad {\rm or}\quad
 D_u^{\pi^+ -\pi^- }(z,Q^2) = \frac{18\left( F_{1}^n\right)_{LO} R_n^{\pi^+
-\pi^-}}{4d_V-u_V}. \label{alpha}
 \end{eqnarray}
SU(2) and $C$ invariance determines the other non-singlet
combinations: \begin{eqnarray} D_d^{\pi^+ -\pi^- }=-D_u^{\pi^+ -\pi^- }, \quad
D_s^{\pi^+ -\pi^- }=D_{\bar s}^{\pi^+ -\pi^- }=0\label{Dh-barh}.
\end{eqnarray} The analogous relations for $D_q^{K^+ -K^- }$ and
$D_q^{\Lambda -\bar\Lambda }$ are given in~\cite{Strategy}.

Via $(K^+,K^-)$ or $(\Lambda,\bar\Lambda )$ production we can
test the conventionally made assumptions
 $s=\bar s$ and $\Delta s=\Delta\bar s$. For the unpolarized case,
 assuming  $D_d^{K^+-K^-}=0$ we have:
 \begin{eqnarray}
(s-\bar s) D_s^{K^+-K^-}& =&
18\,\left(F_1^p\right)_{LO} R_p^{K^+ -K^-} -
 4u_V D_u^{K^+ -K^-}\nonumber\\ &=&18\, \left(F_1^n\right)_{LO}
R_n^{K^+ -K^-} -
 4d_V D_u^{K^+ -K^-}.\label{e1}
 \end{eqnarray}
As for the $K$-meson system $s$ is
a valence quark,  $D_s^{K^+ -K^- }$ should not be a
small quantity. Then a zero value of (\ref{e1}) would imply
$s-\bar s = 0$.

 Having thus determined $(s-\bar s) D_s^{K^+-K^-}$ we can proceed to determine
 $(\Delta s-\Delta\bar s) D_s^{K^+-K^-}$:
\begin{eqnarray}
\Delta A_p^{K^+ -K^-} =
\frac{4\Delta u_V D_u^{K^+ -K^-} +
(\Delta s-\Delta\bar s)D_s^{K^+ -K^-}} {4 u_V D_u^{K^+ -K^-} +
(s-\bar s)D_s^{K^+ -K^-}}\nonumber\\
\Delta A_n^{K^+ -K^-}= \frac{4\Delta d_V D_u^{K^+ -K^-} + (\Delta
s-\Delta\bar s)D_s^{K^+ -K^-}} {4 d_V D_u^{K^+ -K^-} + (s-\bar
s)D_s^{K^+ -K^-}}, \label{e2} \end{eqnarray} $D_u^{K^+-K^-}$ is assumed to
be known through the analogue of (\ref{Dh-barh}). The analogous
relations for $\Lambda,\bar\Lambda$ production can be found
in~\cite{Strategy}

The relations in this paragraph are true in LO approximation only.
A characteristic feature of these expressions, (\ref{A}) and (\ref{alpha}),
is that the RH sides which, in principle, depend on $(x,z,Q^2)$,
should depend only on two of these,
either $(x,Q^2)$  -- eq. (\ref{A}), or $(z,Q^2)$ -- eq.
(\ref{alpha}), so that
 there is an independence of the third variable, which we call
 {\it passive} variable. Each expression should be tested for
 a dependence on the passive variable.

 \vspace{1cm}
 {\bf 4.} The analysis in NLO is more complicated as the simple products
 in the LO expressions are replaced by convolutions.
  This means that instead of solving algebraic equations,
  in NLO we have to deal with integral equations,
  i.e. one needs a fit to the data.

  Using charge conjugation invariance one obtains, for
semi-inclusive pion production
\begin{eqnarray}
&&R_p^{\pi^+-\pi^-} = \frac{[4u_V -d_V][1+\otimes (\alpha_s/2\pi)
{C}_{qq}\otimes ] D_u^{\pi^+ -\pi^-}}{18F_1^p\,[1+2\gamma
(y)\,R^{\,p}]}\nonumber\\
 &&R_n^{\pi^+-\pi^-} =\frac{[4d_V
-u_V][1+\otimes (\alpha_s/2\pi)
{C}_{qq}\otimes ]
D_u^{\pi^+ -\pi^-}}{18F_1^n\,[1+2\gamma
(y)\,R^{\,n}]}.\label{DuNLO}
 \end{eqnarray}
 The only unknown function in
these expressions is $D_u^{\pi^+ -\pi^-}(z,Q^2)$, which evolves as
a non-singlet and does not mix with other FF's. A $\chi^2$
analysis of either or both of the equations (\ref{DuNLO}) should
thus determine
 $D_u^{\pi^+ -\pi^-}$ in NLO without serious ambiguity.

Assuming $D_u^{\pi^+ -\pi^-}$ is now known, one can then
determine $\Delta u_V$ and $\Delta d_V$ in NLO via the equations
\begin{eqnarray}
\Delta A_p^{\pi^+ -\pi^-}
= \frac{ (4\Delta u_V -\Delta d_V)
[1+ \otimes (\alpha_s/2\pi) \Delta C_{qq}\otimes ] D_u^{\pi^+
-\pi^-}} {(4u_V -d_V)[1+\otimes (\alpha_s /2\pi) {
C}_{qq}\otimes ] D_u^{\pi^+ -\pi^-}}\label{valence1NLO}
\end{eqnarray}
\begin{eqnarray}
\Delta A_n^{\pi^+ -\pi^-}
= \frac{ (4\Delta d_V -\Delta u_V)
[1+ \otimes (\alpha_s /2\pi) \Delta C_{qq}\otimes ] D_u^{\pi^+
-\pi^-}} {(4d_V -u_V)[1+\otimes (\alpha_s /2\pi) {
C}_{qq}\otimes ] D_u^{\pi^+ -\pi^-}}\label{valence2NLO}
\end{eqnarray}
where, of course, $\Delta u_V$ and $\Delta d_V$ evolve as
non-singlets and do
 not mix with other densities. Eqs. (\ref{valence1NLO}) and
 (\ref{valence2NLO}) determine the densities $\Delta u_V$ and
 $\Delta d_V$ in NLO without any assumptions about the less known
 polarized gluon and sea densities.

 Once $\Delta u_V$ and $\Delta d_V$ are known in NLO we can
calculate $[\Delta\bar u(x,Q^2) - \Delta\bar d(x,Q^2)]_{NLO}$

from (\ref{sea}), using the NLO expression (\ref{deltaq3})
for $\Delta q_3$.

The expressions for $s-\bar s$ and $\Delta s-\Delta\bar s$ in NLO
can be found in~\cite{Strategy}.

\vspace{1cm} {\bf 5.} Recently\begin{wrapfigure}{-l}{9cm}
\epsfig{figure=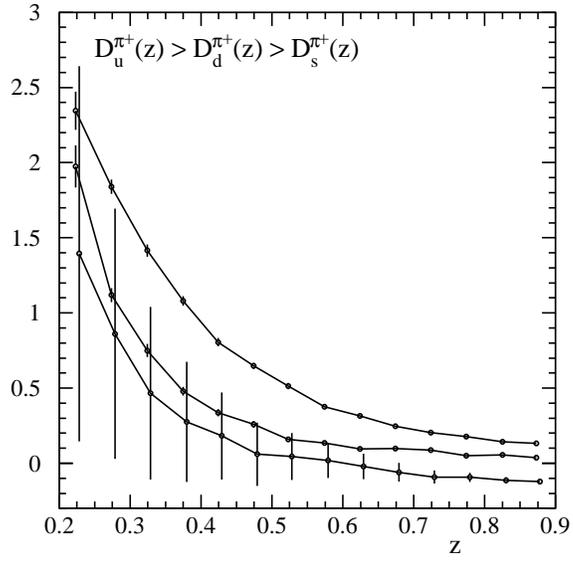,width=9cm}
\vspace{-0.5cm}
\caption{$D_u^{\pi^+}$, $ D_d^{\pi^+}$ and $D_s^{\pi^+}$
extracted as explained in the text.}
\end{wrapfigure} 
the HERMES group has  published
very precise  data for  unpolarized SIDIS for $\pi^\pm$
production~\cite{HERMES}. There are 3  independent FFs
$D_u^{\pi^+}$, $ D_d^{\pi^+}$ and $D_s^{\pi^+}$.  The SIDIS data
provide two equations for them, and a third piece of information
is thus required. We have shown~\cite{FFs} that, though the
individual flavoured $D_q^{\pi^+}$ in $e^+e^-\to \pi^\pm +X$ are
poorly known, the  singlet combination $D_\Sigma^{\pi^+} =
2\left(D_u^{\pi^+}+D_d^{\pi^+}+D_s^{\pi^+}\right)$ is very well
 determined  at the $Z^0$
peak. Then, evolving this down to $Q^2$ of the SIDIS experiment,
it allows us  to extract $D_u^{\pi^+}$, $ D_d^{\pi^+}$ and
$D_s^{\pi^+}$  without any  assumptions about favoured and
unfavoured transitions. The results are given in the Figure. The
uncertainties, mainly affecting $D_S^{\pi^+}$, come from the
evolution of $D_\Sigma^{\pi^+}$ at $Q^2 \approx m_Z^2$ down to
$D_\Sigma^{\pi^+}$ at $Q^2\approx few (GeV)^2$, as it involves
mixing with the poorly known $D_G^{\pi^+}$. Both $D_u^{\pi^+}$ and
$ D_d^{\pi^+}$ are quite accurately determined. Quite surprising
is the magnitude of the "unfavoured" fragmentation $d\to \pi^+$.

\end{document}